\documentclass[%
 reprint,
 twocolumn,
 superscriptaddress,
 amsmath,amssymb,
 aps,
]{revtex4-1}

\usepackage{dcolumn}
\usepackage{bm}
\usepackage[utf8]{inputenc}
\usepackage[T1]{fontenc}
\usepackage{graphicx}
\usepackage{tabularx}

\usepackage{hyperref}
\begin{document}
\title{Inorganic Materials Synthesis Planning with Literature-Trained Neural Networks}

\author{Edward Kim}
\affiliation{Dept. of Materials Science and Engineering, Massachusetts Institute of Technology, Cambridge, MA, USA}
\author{Zach Jensen}
\affiliation{Dept. of Materials Science and Engineering, Massachusetts Institute of Technology, Cambridge, MA, USA}
\author{Alexander van Grootel}
\affiliation{Dept. of Materials Science and Engineering, Massachusetts Institute of Technology, Cambridge, MA, USA}
\author{Kevin Huang}
\affiliation{Dept. of Materials Science and Engineering, Massachusetts Institute of Technology, Cambridge, MA, USA}
\author{Matthew Staib}
\affiliation{Dept. of EECS and CSAIL, Massachusetts Institute of Technology, Cambridge, MA, USA}
\author{Sheshera Mysore}
\affiliation{College of Information and Computer Sciences, University of Massachusetts Amherst, Amherst, MA, USA}
\author{Haw-Shiuan Chang}
\affiliation{College of Information and Computer Sciences, University of Massachusetts Amherst, Amherst, MA, USA}
\author{Emma Strubell}
\affiliation{College of Information and Computer Sciences, University of Massachusetts Amherst, Amherst, MA, USA}
\author{Andrew McCallum}
\affiliation{College of Information and Computer Sciences, University of Massachusetts Amherst, Amherst, MA, USA}
\author{Stefanie Jegelka}
\affiliation{Dept. of EECS and CSAIL, Massachusetts Institute of Technology, Cambridge, MA, USA}

\author{Elsa Olivetti}
\email{Corresponding author: elsao@mit.edu}
\affiliation{Dept. of Materials Science and Engineering, Massachusetts Institute of Technology, Cambridge, MA, USA}


\date{February 17, 2019}

\begin{abstract}
Leveraging new data sources is a key step in accelerating the pace of materials design and discovery. To complement the strides in synthesis planning driven by historical, experimental, and computed data, we present an automated method for connecting scientific literature to synthesis insights. Starting from natural language text, we apply word embeddings from language models, which are fed into a named entity recognition model, upon which a conditional variational autoencoder is trained to generate syntheses for arbitrary materials. We show the potential of this technique by predicting precursors for two perovskite materials, using only training data published over a decade prior to their first reported syntheses. We demonstrate that the model learns representations of materials corresponding to synthesis-related properties, and that the model's behavior complements existing thermodynamic knowledge. Finally, we apply the model to perform synthesizability screening for proposed novel perovskite compounds.
\end{abstract}

\flushbottom
\maketitle

\thispagestyle{empty}

Recent advances in predicting material properties \cite{xie2018crystal,huan2015accelerated, isayev2015materials}, screening synthesizable compounds \cite{meredig2014combinatorial, aykol2018network, kim2018machine}, and organic reaction prediction \cite{segler2018planning, gao2018using, coley2017prediction} have been driven, in part, by the accessibility of machine-readable datasets \cite{saal2013materials, jain2013commentary, goodman2009computer} and, consequently, data-driven models. In stark contrast to organic reaction databases \cite{goodman2009computer}, the overwhelming majority of \emph{inorganic} synthesis knowledge lies locked within the text of journal articles \cite{kim2017materials} and laboratory notebooks \cite{raccuglia2016machine}. While the latter has been shown as an effective source for guiding successful syntheses, there is currently no general framework for automatically drawing synthesis insights from the literature at large.

Although scientific literature has previously been used to illuminate patterns in nanoscale morphologies \cite{kim2017materials}, device performances \cite{ghadbeigi2015performance}, and apparatus parameters \cite{young2018data}, each of these efforts have required tailored data representations and algorithms. To rapidly translate literature knowledge into synthesis planning resources without fine-tuned adjustments, transfer learning across different materials systems is necessary \cite{kim2017materials}. Thus, leveraging data from broad volumes of scientific literature is a critical step towards capturing synthesis trends and extending them to unknown materials.

In this work, we present an automated method for connecting scientific literature to insights for materials synthesis planning. We show that an unsupervised conditional variational autoencoder (CVAE) \cite{kingma2013auto,sohn2015learning,gomez2018automatic} can generate synthesis predictions for a variety of materials, including materials unknown to the model. This CVAE learns directly from the materials synthesis literature and produces an internal representation of precursors which corresponds to physical and chemical trends without receiving any explicit domain knowledge. We use the literature knowledge captured by the CVAE to complement first-principles techniques in materials screening tasks.

To accelerate the efforts of the materials science community, we open-source several key resources used in this work \footnote{\url{www.github.com/olivettigroup/materials-synthesis-generative-models}}: We release context-sensitive embeddings from language models (ELMo) that have been adapted for materials science text \cite{peters2018deep} along with a pre-trained FastText word embedding model for materials science \cite{bojanowski2016enriching}. Each of these embedding models has been trained on a collection of over 2.5 million materials science journal articles \cite{kim2017materials, kim2017machine}. Finally, we provide over two hundred annotated literature synthesis routes for named entity recognition (NER) tasks, such as identifying reaction conditions and materials.

\begin{figure}[htbp]
\centering
\includegraphics[width=\linewidth]{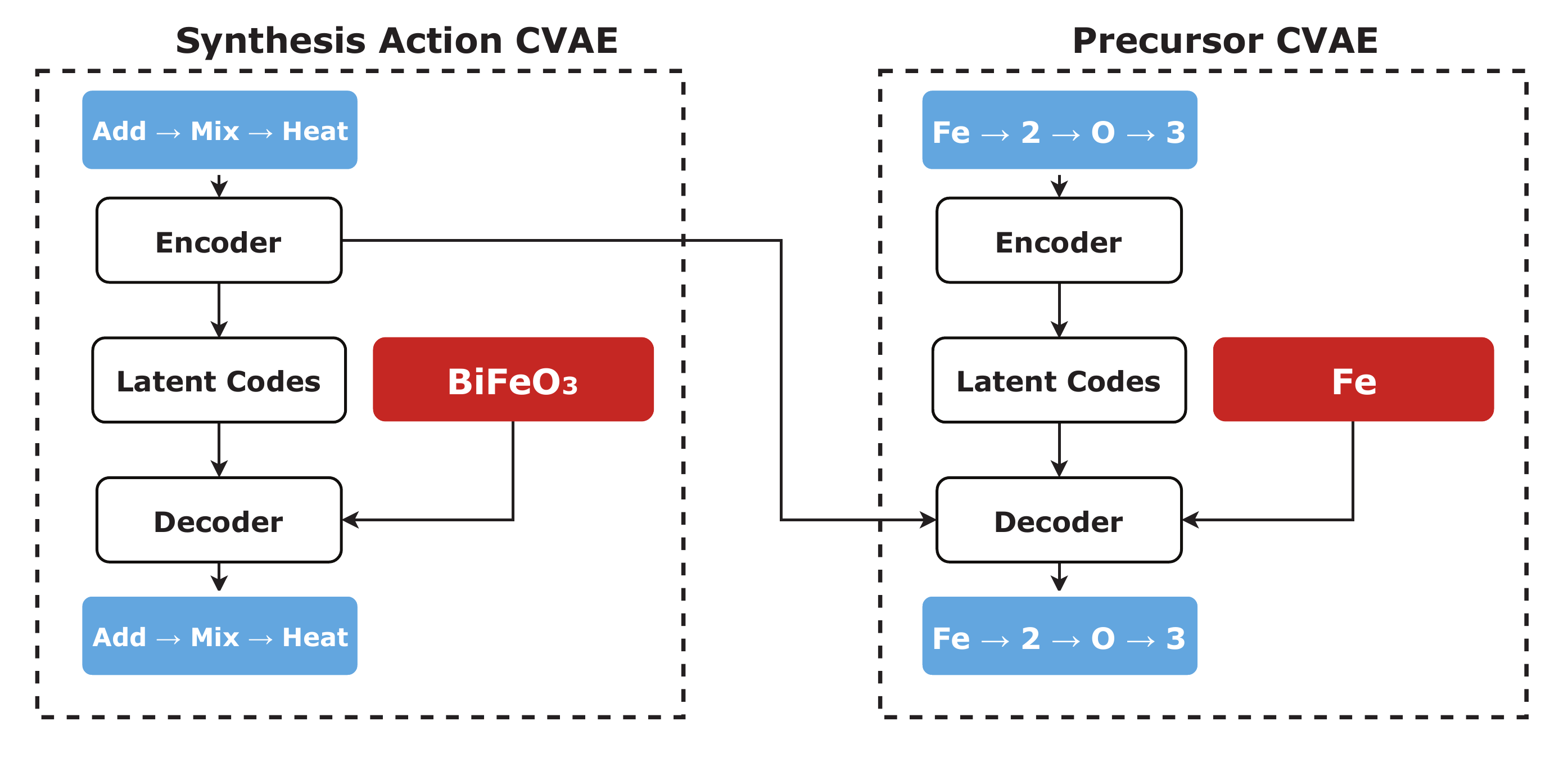}
\caption{Schematic diagram of the CVAE architecture. The model consists of two joined CVAEs used for learning synthesis actions and precursors, respectively. The Synthesis Action CVAE learns distributions of synthesis actions sequences conditioned on target materials. The Precursor CVAE learns distributions of precursor formulas conditioned on both a target element and an encoded representation of the jointly-observed synthesis action sequence. Target materials are represented by FastText embeddings, and all other inputs to the model are sequences of one-hot vectors.}
\label{fig:cvae_schematic}
\end{figure}

We first describe our automated workflow. After a recurrent neural network \cite{chung2014empirical} identifies synthesis sections of journal articles, context-sensitive ELMo word embeddings are computed and passed into another recurrent neural network which performs NER to identify precursors, synthesis target materials, and synthesis actions. Then, a CVAE model, shown in Figure \ref{fig:cvae_schematic}, is trained to learn representations of synthesis routes from the named entities in an unsupervised manner. More details on these methods are provided in the Supplementary Methods.

To maximize the opportunity for transfer learning of synthesis trends, we choose a broad definition for synthesis routes that requires minimal assumptions. For a given target material $m$, a synthesis route $S_{m}^i$ is a 2-tuple consisting of a sequence of $n$ synthesis actions $(a_1, a_2, \ldots, a_n)$ acting on a set of $l$ precursors $\{p_1, \ldots, p_l\}$,
\begin{equation}
    S_{m}^i = \left((a_k)^n, \{p_j\}^l\right)
\end{equation}

\noindent and in general, a single target material $m$ may have $N > 1$ valid synthesis routes and thus $S_{m}^i$ represents the $i$th valid synthesis route for $m$. We also define precursors $p$ as ``element sources,'' such that they are materials sharing an element with $m$. The CVAE model is then constructed to model the following distributions,
\begin{align}
    \mathbb{P}&((a_k)^n|\theta_a, m) \\
    \mathbb{P}&\left(p_j|\theta_p, e_j, (a_k)^n \right)
\end{align}

\noindent where $\theta_a$ and $\theta_p$ are model parameters for the synthesis action and precursor CVAEs, respectively,
and $e_j$ is the shared element between a precursor $p_j$ and target material $m$ (e.g., titanium).

Since CVAEs are generative models, novel synthesis actions and precursors can be generated by sampling from a Gaussian prior distribution \cite{kingma2013auto}. Critically, we represent $m$ by FastText word embeddings which enable the transfer of synthesis trends between existing and novel materials by leveraging literature-based similarity.

\begin{table}[!htbp]
\caption{Generated precursors for InWO$_3$ and PbMoO$_3$, drawn from the CVAE model. The CVAE model was trained on synthesis routes published during or before 2005. A more detailed version is available in the Supplemental Results.}
\begin{tabular}{l r} 
 \centering
 Target Material & Precursors \\ 
 [0.5ex] 
 \hline\hline
  & In$_2$S$_3$ + WCl$_4$ \\ 
  & In(NO$_3$)$_3$ + WCl$_4$ \\
 InWO$_3$ &  In$_2$O$_3$ + WO$_2$ \\
   & In$_2$O$_3$ + WN  \\
  & $^\dagger$InCl$_3$ + Na$_2$WO$_4$\\ 
 [0.5ex]
 \hline
  & PbCl$_2$ + MoCl$_2$ \\ 
 PbMoO$_3$ & PbSO$_4$ + MoCl$_2$  \\ 
  & $^\ddagger$PbO + MoO$_2$ \\ 
 [0.5ex] 
 \hline\hline
 \multicolumn{2}{l}{
  \begin{minipage}{8cm}
  \vspace{0.2cm}
  \raggedright
    $^\dagger$ Precursors match Kamalakkannan et al. (2016) \cite{kamalakkannan2016synthesis}. \\
     $^\ddagger$ Precursors match Takatsu et al. (2017) \cite{takatsu2017cubic}.
  \end{minipage}
 }\\
\end{tabular}
\\
\label{table:2005_generated_recipes}
\end{table}

To demonstrate the applicability of our CVAE method, we construct a dataset of approximately 51,000 synthesis action sequences and 116,000 precursors via a general set of search terms (``perovskite + thermoelectric + multiferroic + photovoltaic + solar + nano + cathode'') and apply our neural network pipeline. We investigate the effectiveness of the CVAE model in synthesis planning by performing a publication-year-split experiment, where the model is trained only on syntheses published prior to 2005 ($\sim$2800 syntheses). We apply the model in predicting precursors for materials that were unseen during training, are computationally predicted as stable perovskites \cite{xie2018crystal}, and only recently appear in the literature: InWO$_3$ and PbMoO$_3$, first reported in 2016 and 2017, respectively \cite{kamalakkannan2016synthesis,takatsu2017cubic}. Table \ref{table:2005_generated_recipes} shows a report of the data generated by sampling from the CVAE's Gaussian prior distributions, where the CVAE suggests the precursors for both materials (see Table \ref{table:supp_2005_generated_recipes} for additional details). The CVAE model is thus capable of predicting synthesis precursors while relying only on literature knowledge from more than a decade prior to the literature-reported syntheses of these materials. Trial-and-error (or random) precursor selection is substantially less efficient, as the number of possible precursor sets for each material is in the hundreds. Thus literature-driven models may greatly accelerate future synthesis attempts of novel materials.

During the data generation process, the CVAE model proposes several plausible syntheses beyond the literature-matching samples. To the best of the authors' knowledge, the only reported synthesis of InWO$_3$ is via a solution-phase route. However, the CVAE model suggests that solid state synthesis of InWO$_3$ may be possible, using In$_2$O$_3$ and either WO$_2$ or WN as precursors. Such syntheses may be feasible, as they are thermodynamically favorable (using data at 0 K and 0 atm from OQMD) \cite{saal2013materials}:
\begin{align*}
    \text{In}_2\text{O}_3 + 2\text{WO}_2 &\longrightarrow 2\text{InWO}_3 + 2\text{O}_2 \\
    \Delta \text{H} &= -158 \text{ kJ/mol} \\
    \\
    \text{In}_2\text{O}_3 + 2\text{WN} &\longrightarrow 2\text{InWO}_3 + \text{N}_2 \\
    \Delta \text{H} &= -930 \text{ kJ/mol}
\end{align*}

\noindent We do note, however, that these simple thermodynamic analyses can only be used as rough guidelines. Besides the limitations of estimating an overall thermochemical reaction for the synthesis, along with extrapolation from STP conditions, kinetic effects are not considered here. Indeed, while it is common to mix binary oxide precursors in solid state syntheses of ternary (or quaternary, etc.) oxides, the use of nitride precursors is far less common due to the high bond energies of many nitride compounds \cite{sun2017thermodynamic}. To achieve a clearer understanding of kinetic effects, experimental verification would be required alongside a model which incorporates reaction conditions (e.g., temperatures), and this is an area for future work.

In the suggested recipes for PbMoO$_3$, the CVAE model suggests a solution-phase route using PbSO$_4$ and MoCl$_2$, both of which are soluble under acidic conditions. This may provide another viable path towards synthesizing PbMoO$_3$, which has only been realized so far by solid state synthesis methods \cite{takatsu2017cubic}. Thus, the CVAE may be used as a source of suggestions for synthesis planning, although we stress that these suggestions still need human evaluation, further analysis, and cannot be applied ``out of the box.''

Despite the fact that chemical knowledge is never given to the CVAE model, solubility rules emerge from the model results. To demonstrate this, we generate W-bearing precursors for InWO$_3$ conditioned on two representative action sequences sampled from the Synthesis Action CVAE: A solid-state synthesis (mix, grind, calcine, press, sinter, cool) and a solution-phase synthesis (add, dissolve, stir, heat, wash, dry). Following this, we generate 10,000 CVAE-suggested precursors. The most common W-bearing precursor generated for the solid-state synthesis is the water-insoluble WO$_3$, while the the most common precursor generated for the solution-phase synthesis is the highly-soluble Na$_2$WO$_4$. The differences of these precursor likelihoods in each case is substantial, with $-16\%$ and $+21\%$ changes to the likelihoods of the CVAE suggesting WO$_3$ and Na$_2$WO$_4$, respectively, when switching from conditioning on solid-state to solution-phase synthesis actions. 

This effect of learning precursor trends from the literature is further demonstrated upon inspecting latent codes learned by the model. Since the CVAE learns conditional distributions, the input precursors are projected into a degenerate latent space, where the degeneracy is split by the conditional input received by the decoder. By investigating several examples (see Figure \ref{fig:supp_latent_degen}), we find that the CVAE learns to group precursors with similar synthesis-relevant properties, including insoluble binary oxides, water-soluble polyanion compounds, and  pure/alloyed metals. This suggests that the CVAE model is capable of capturing chemical intuition and composition-driven similarity solely by joint observations of precursors, synthesis actions, and target materials. Despite the lack of ``negative'' data in the literature, the diversity of published synthesis literature is sufficient to drive the CVAE model in learning meaningful representations of precursors.

\begin{figure*}[!htbp]
\centering
\includegraphics[width=\linewidth]{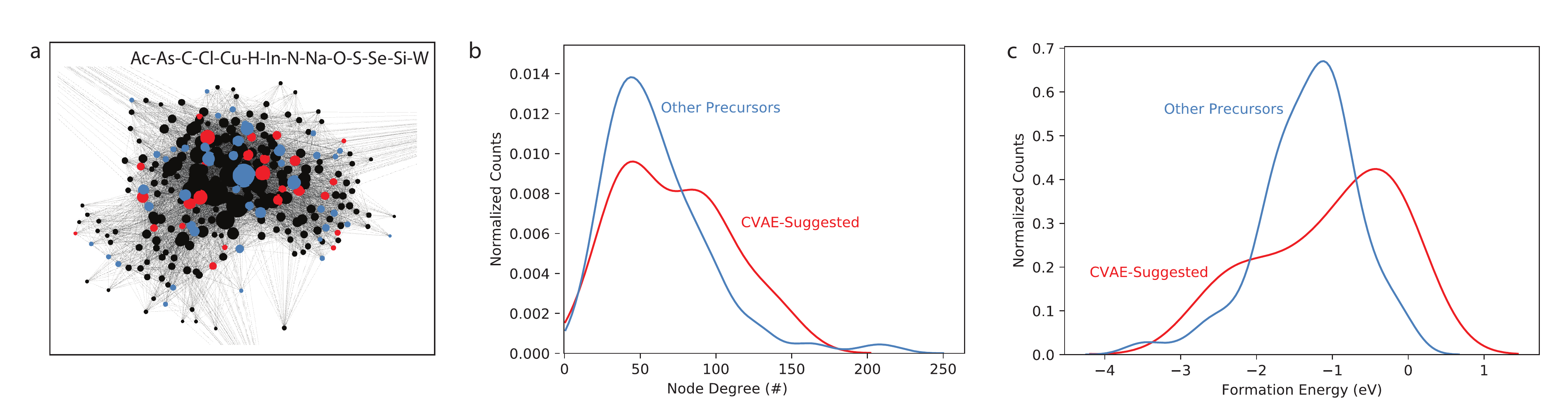}
\caption{Graph representation of a phase diagram for InWO$_3$ precursor chemical space \cite{hegde2018phase}, using data from 1000 precursor sets generated by the CVAE. a) Zoomed-in view of phase diagram graph for the 14-element chemical system. Red (n = 27) and blue (n = 46) colored nodes are precursors for InWO$_3$ (i.e., materials containing In or W), and all other material nodes are black (n = 198). Red nodes  are suggested at least once by the CVAE, and blue nodes are never suggested by the CVAE. Graph edges represent two-phase equilibria, as determined by OQMD \cite{saal2013materials, hegde2018phase}, and node sizes are proportional to node degree. b) Distributions of graph node degrees (i.e., number of two-phase equilibria) for CVAE-suggested and non-suggested precursors. c) Distributions of formation energies for CVAE-suggested and non-suggested precursors.}
\label{fig:precursors}
\end{figure*}

To emphasize the particular nature of synthesis planning via a \textit{literature-trained} model, we contrast suggested precursors by the CVAE model with thermodynamic stability computations from OQMD \cite{saal2013materials}. Figure \ref{fig:precursors}a shows a graph representation of the chemical space spanned by all CVAE-suggested precursors, where graph vertices are precursors and graph edges represent thermodynamic two-phase equilibria \cite{hegde2018phase}. Clearly, the CVAE is not simply suggesting all thermodynamically-viable precursors, as there is a substantial set of precursors never suggested by the CVAE (blue vertices versus red vertices). Indeed, the CVAE has filtered precursors from a set of 73 possible precursors to only 27. Additionally, Figures \ref{fig:precursors}b and \ref{fig:precursors}c show that the CVAE is not selecting precursors in correspondence with isolated thermodynamic metrics: the CVAE's suggestions are explained neither by thermodynamic reactivity (i.e., the number of relevant two-phase equilibria with respect to other precursors) nor individual precursor stability (i.e., formation energy). This suggests that there is a meaningful difference between the thermodynamically-driven and literature-driven synthesis planning methods. While the former probes the realm of physical possibility, the latter emphasizes practical choices and historical trends.

We next train the CVAE model on our full dataset, using no publication-year cutoffs. To investigate the capability of the model for suggesting syntheses of a novel, never-before-synthesized material, we consider novel ABO$_3$ perovskite materials proposed by Balachandran et al. \cite{balachandran2018predictions}.  These proposed perovskites have not previously been synthesized, and have high thermodynamic stability as measured by energy differences against their convex hulls. We note that ABO$_3$ perovskites are used here as a representative example due to their chemical variety and diverse range of properties, but the CVAE model does indeed generalize to other categories of materials (see Table \ref{table:supp_alt_candidates}).

HgZrO$_3$ is one such example of a thermodynamically stable, unsynthesized perovskite material  \cite{balachandran2018predictions}, and we perform synthesis predictions using the CVAE model (see Table \ref{table:supp_generated_unseen}). We find that the CVAE proposes solid-state syntheses which appear to be thermodynamically reasonable:
\begin{align*}
    \text{HgO} + \text{ZrC} + 2\text{O}_2 &\longrightarrow \text{HgZrO}_3 + \text{CO}_2 \\
    \Delta \text{H} &= -1340 \text{ kJ/mol} \\
    \\
    \text{HgO} + \text{ZrO}_2 &\longrightarrow \text{HgZrO}_3 \\
    \Delta \text{H} &= -0.29 \text{ kJ/mol}
\end{align*}

\noindent Again, we emphasize that thermodynamic analyses are often insufficient to evaluate reaction plausibility. As an additional utility for evaluating generated synthesis parameters, we develop a similarity metric based on the latent codes learned by the CVAE. By measuring nearest-neighbors of latent codes for the recipe using mercuric oxide and zirconium carbide, we find that the two closest literature recipes are for solid-state syntheses of SrZrO$_3$ and BaAl$_2$O$_4$ (see Figure \ref{fig:supp_latent_nns}). Besides providing insight into which observed literature examples ``inspired'' this particular prediction, we are also led to further insights on precursor selections. ZrC is an uncommon choice of precursor, but carbonate precursors are readily used in solid-state syntheses. Indeed, both of the near-neighbor syntheses for HgZrO$_3$ use carbonate precursors rather than carbides.

While similarity methods have previously been produced for materials (e.g., based on crystal structures) \cite{yang2013data}, the CVAE incorporates synthesis knowledge to produce a distinct measure of similarity. Indeed, from a structural point of view, it would not be expected that SrZrO$_3$ and BaAl$_2$O$_4$ should have high similarity to HgZrO$_3$, since all three materials form ground-state structures in different crystal systems (orthorhombic, hexagonal, and cubic, respectively).

Moreover, rather than measuring similarity against entire materials, the CVAE-based metric operates at the level of individual reported (or generated) synthesis routes. Since nearest-neighbor search is computationally inefficient in high dimensional spaces, the dimensionality reduction imposed by the CVAE enables this ``latent citation'' model to be used as a rapid, data-driven synthesis planning method.

Finally, we present results for synthesis screening using the CVAE model to suggest syntheses for numerous ABO$_3$ suggested by Balachandran et al. \cite{balachandran2018predictions}. The CVAE was used to generate syntheses with ten data generation attempts per compound, and only compounds which had at least one suggested synthesis route with commercially-available precursors were considered to have passed the test, which is the same criterion used by Segler et al. \cite{segler2018planning} to evaluate retrosynthetic routes for organic molecules.

\begin{figure}[!htbp]
\centering
\includegraphics[width=\linewidth]{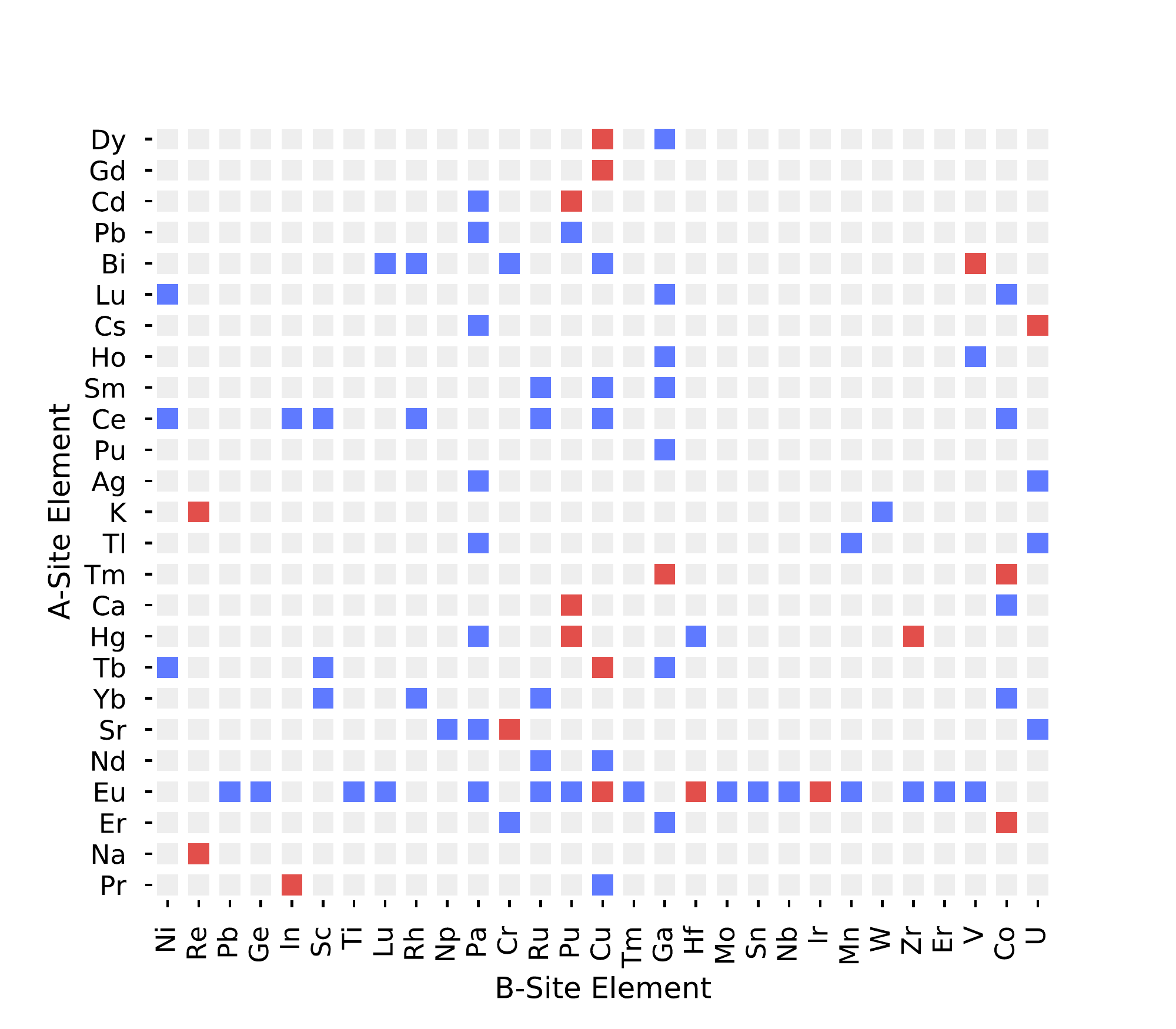}
\caption{Unsynthesized ABO$_3$ perovskite compounds, labelled by their A-site and B-site elements. Colored-in squares are perovskites predicted to be stable \cite{balachandran2018predictions}. Red and blue colors correspond to compounds which passed or failed the CVAE screening, respectively.}
\label{fig:unsynth_heatmap}
\end{figure}

Figure \ref{fig:unsynth_heatmap} shows a grid of possible A-site and B-site atoms for ABO$_3$ perovskite materials, with screened compounds represented by highlighted combinations of A-site and B-site atoms. While a joint machine learning and density functional theory method \cite{balachandran2018predictions} selects a set of materials which are thermodynamically stable in the perovskite form, the further-imposed synthesis screening selects a subset that is most readily synthesizable based on existing literature knowledge. From a set of 83 proposed ABO$_3$ perovskite compounds, the CVAE has selected a subset of only 19.

The CVAE model, combined with the rest of our neural network workflow, enables a new axis of synthesis screening which complements existing domain knowledge \cite{balachandran2018predictions, aykol2018network}. By incorporating and extending patterns in historical literature, materials which are theoretically synthesizable can be rapidly and automatically filtered by their practical synthesizability. While the methods presented in this paper are applicable to various materials systems and synthesis methods, we recognize that our broad representation of synthesis routes omits information such as temperatures, solvents, and morphologies, and additionally assumes that there is a one-to-one relation between elements in precursors and targets. We thus believe that promising future work lies in the direction of generative models with narrower scope, but finer-grained detail. For example, limiting the dataset to solvothermal syntheses may facilitate prediction of solvent choices, solvothermal reaction temperatures, and dwell times. This additional domain knowledge may be incorporated by filtering proposed synthesis parameters \cite{segler2018planning} or constraining model outputs \cite{kusner2017grammar}. Motivated by these possibilities, our open-source NER annotations include the necessary labels (e.g., reaction conditions) to enable these future studies.

We would like to acknowledge funding from the National Science Foundation Award 1534340, DMREF that provided support to make this work possible, support from the Office of Naval Research (ONR) under Contract No. N00014-16-1-2432, the MIT Energy Initiative, and NSF CAREER \#1553284. Early work was collaborative under the Dept. of Energy’s Basic Energy Science Program through the Materials Project under Grant No. EDCBEE. This work was also partly funded by the MIT-Sensetime Alliance on Artificial Intelligence. We would also like to acknowledge valuable feedback from Rafael Jaramillo, Gerbrand Ceder, Olga Kononova, Wenhao Sun, Haoyan Huo, and Tanjin He.


\bibliographystyle{apsrev4-1}
\bibliography{main}

\begin{thebibliography}{37}%
\makeatletter
\providecommand \@ifxundefined [1]{%
 \@ifx{#1\undefined}
}%
\providecommand \@ifnum [1]{%
 \ifnum #1\expandafter \@firstoftwo
 \else \expandafter \@secondoftwo
 \fi
}%
\providecommand \@ifx [1]{%
 \ifx #1\expandafter \@firstoftwo
 \else \expandafter \@secondoftwo
 \fi
}%
\providecommand \natexlab [1]{#1}%
\providecommand \enquote  [1]{``#1''}%
\providecommand \bibnamefont  [1]{#1}%
\providecommand \bibfnamefont [1]{#1}%
\providecommand \citenamefont [1]{#1}%
\providecommand \href@noop [0]{\@secondoftwo}%
\providecommand \href [0]{\begingroup \@sanitize@url \@href}%
\providecommand \@href[1]{\@@startlink{#1}\@@href}%
\providecommand \@@href[1]{\endgroup#1\@@endlink}%
\providecommand \@sanitize@url [0]{\catcode `\\12\catcode `\$12\catcode
  `\&12\catcode `\#12\catcode `\^12\catcode `\_12\catcode `\%12\relax}%
\providecommand \@@startlink[1]{}%
\providecommand \@@endlink[0]{}%
\providecommand \url  [0]{\begingroup\@sanitize@url \@url }%
\providecommand \@url [1]{\endgroup\@href {#1}{\urlprefix }}%
\providecommand \urlprefix  [0]{URL }%
\providecommand \Eprint [0]{\href }%
\providecommand \doibase [0]{http://dx.doi.org/}%
\providecommand \selectlanguage [0]{\@gobble}%
\providecommand \bibinfo  [0]{\@secondoftwo}%
\providecommand \bibfield  [0]{\@secondoftwo}%
\providecommand \translation [1]{[#1]}%
\providecommand \BibitemOpen [0]{}%
\providecommand \bibitemStop [0]{}%
\providecommand \bibitemNoStop [0]{.\EOS\space}%
\providecommand \EOS [0]{\spacefactor3000\relax}%
\providecommand \BibitemShut  [1]{\csname bibitem#1\endcsname}%
\let\auto@bib@innerbib\@empty
\bibitem [{\citenamefont {Xie}\ and\ \citenamefont
  {Grossman}(2018)}]{xie2018crystal}%
  \BibitemOpen
  \bibfield  {author} {\bibinfo {author} {\bibfnamefont {T.}~\bibnamefont
  {Xie}}\ and\ \bibinfo {author} {\bibfnamefont {J.~C.}\ \bibnamefont
  {Grossman}},\ }\href@noop {} {\bibfield  {journal} {\bibinfo  {journal}
  {Physical Review Letters}\ }\textbf {\bibinfo {volume} {120}},\ \bibinfo
  {pages} {145301} (\bibinfo {year} {2018})}\BibitemShut {NoStop}%
\bibitem [{\citenamefont {Huan}\ \emph {et~al.}(2015)\citenamefont {Huan},
  \citenamefont {Mannodi-Kanakkithodi},\ and\ \citenamefont
  {Ramprasad}}]{huan2015accelerated}%
  \BibitemOpen
  \bibfield  {author} {\bibinfo {author} {\bibfnamefont {T.~D.}\ \bibnamefont
  {Huan}}, \bibinfo {author} {\bibfnamefont {A.}~\bibnamefont
  {Mannodi-Kanakkithodi}}, \ and\ \bibinfo {author} {\bibfnamefont
  {R.}~\bibnamefont {Ramprasad}},\ }\href@noop {} {\bibfield  {journal}
  {\bibinfo  {journal} {Physical Review B}\ }\textbf {\bibinfo {volume} {92}},\
  \bibinfo {pages} {014106} (\bibinfo {year} {2015})}\BibitemShut {NoStop}%
\bibitem [{\citenamefont {Isayev}\ \emph {et~al.}(2015)\citenamefont {Isayev},
  \citenamefont {Fourches}, \citenamefont {Muratov}, \citenamefont {Oses},
  \citenamefont {Rasch}, \citenamefont {Tropsha},\ and\ \citenamefont
  {Curtarolo}}]{isayev2015materials}%
  \BibitemOpen
  \bibfield  {author} {\bibinfo {author} {\bibfnamefont {O.}~\bibnamefont
  {Isayev}}, \bibinfo {author} {\bibfnamefont {D.}~\bibnamefont {Fourches}},
  \bibinfo {author} {\bibfnamefont {E.~N.}\ \bibnamefont {Muratov}}, \bibinfo
  {author} {\bibfnamefont {C.}~\bibnamefont {Oses}}, \bibinfo {author}
  {\bibfnamefont {K.}~\bibnamefont {Rasch}}, \bibinfo {author} {\bibfnamefont
  {A.}~\bibnamefont {Tropsha}}, \ and\ \bibinfo {author} {\bibfnamefont
  {S.}~\bibnamefont {Curtarolo}},\ }\href@noop {} {\bibfield  {journal}
  {\bibinfo  {journal} {Chemistry of Materials}\ }\textbf {\bibinfo {volume}
  {27}},\ \bibinfo {pages} {735} (\bibinfo {year} {2015})}\BibitemShut
  {NoStop}%
\bibitem [{\citenamefont {Meredig}\ \emph {et~al.}(2014)\citenamefont
  {Meredig}, \citenamefont {Agrawal}, \citenamefont {Kirklin}, \citenamefont
  {Saal}, \citenamefont {Doak}, \citenamefont {Thompson}, \citenamefont
  {Zhang}, \citenamefont {Choudhary},\ and\ \citenamefont
  {Wolverton}}]{meredig2014combinatorial}%
  \BibitemOpen
  \bibfield  {author} {\bibinfo {author} {\bibfnamefont {B.}~\bibnamefont
  {Meredig}}, \bibinfo {author} {\bibfnamefont {A.}~\bibnamefont {Agrawal}},
  \bibinfo {author} {\bibfnamefont {S.}~\bibnamefont {Kirklin}}, \bibinfo
  {author} {\bibfnamefont {J.~E.}\ \bibnamefont {Saal}}, \bibinfo {author}
  {\bibfnamefont {J.}~\bibnamefont {Doak}}, \bibinfo {author} {\bibfnamefont
  {A.}~\bibnamefont {Thompson}}, \bibinfo {author} {\bibfnamefont
  {K.}~\bibnamefont {Zhang}}, \bibinfo {author} {\bibfnamefont
  {A.}~\bibnamefont {Choudhary}}, \ and\ \bibinfo {author} {\bibfnamefont
  {C.}~\bibnamefont {Wolverton}},\ }\href@noop {} {\bibfield  {journal}
  {\bibinfo  {journal} {Physical Review B}\ }\textbf {\bibinfo {volume} {89}},\
  \bibinfo {pages} {094104} (\bibinfo {year} {2014})}\BibitemShut {NoStop}%
\bibitem [{\citenamefont {Aykol}\ \emph {et~al.}(2018)\citenamefont {Aykol},
  \citenamefont {Hegde}, \citenamefont {Suram}, \citenamefont {Hung},
  \citenamefont {Herring}, \citenamefont {Wolverton},\ and\ \citenamefont
  {Hummelsh{\o}j}}]{aykol2018network}%
  \BibitemOpen
  \bibfield  {author} {\bibinfo {author} {\bibfnamefont {M.}~\bibnamefont
  {Aykol}}, \bibinfo {author} {\bibfnamefont {V.~I.}\ \bibnamefont {Hegde}},
  \bibinfo {author} {\bibfnamefont {S.}~\bibnamefont {Suram}}, \bibinfo
  {author} {\bibfnamefont {L.}~\bibnamefont {Hung}}, \bibinfo {author}
  {\bibfnamefont {P.}~\bibnamefont {Herring}}, \bibinfo {author} {\bibfnamefont
  {C.}~\bibnamefont {Wolverton}}, \ and\ \bibinfo {author} {\bibfnamefont
  {J.~S.}\ \bibnamefont {Hummelsh{\o}j}},\ }\href@noop {} {\bibfield  {journal}
  {\bibinfo  {journal} {arXiv preprint arXiv:1806.05772}\ } (\bibinfo {year}
  {2018})}\BibitemShut {NoStop}%
\bibitem [{\citenamefont {Kim}\ \emph {et~al.}(2018)\citenamefont {Kim},
  \citenamefont {Ward}, \citenamefont {He}, \citenamefont {Krishna},
  \citenamefont {Agrawal},\ and\ \citenamefont {Wolverton}}]{kim2018machine}%
  \BibitemOpen
  \bibfield  {author} {\bibinfo {author} {\bibfnamefont {K.}~\bibnamefont
  {Kim}}, \bibinfo {author} {\bibfnamefont {L.}~\bibnamefont {Ward}}, \bibinfo
  {author} {\bibfnamefont {J.}~\bibnamefont {He}}, \bibinfo {author}
  {\bibfnamefont {A.}~\bibnamefont {Krishna}}, \bibinfo {author} {\bibfnamefont
  {A.}~\bibnamefont {Agrawal}}, \ and\ \bibinfo {author} {\bibfnamefont
  {C.}~\bibnamefont {Wolverton}},\ }\href@noop {} {\bibfield  {journal}
  {\bibinfo  {journal} {Physical Review Materials}\ }\textbf {\bibinfo {volume}
  {2}},\ \bibinfo {pages} {123801} (\bibinfo {year} {2018})}\BibitemShut
  {NoStop}%
\bibitem [{\citenamefont {Segler}\ \emph {et~al.}(2018)\citenamefont {Segler},
  \citenamefont {Preuss},\ and\ \citenamefont {Waller}}]{segler2018planning}%
  \BibitemOpen
  \bibfield  {author} {\bibinfo {author} {\bibfnamefont {M.~H.}\ \bibnamefont
  {Segler}}, \bibinfo {author} {\bibfnamefont {M.}~\bibnamefont {Preuss}}, \
  and\ \bibinfo {author} {\bibfnamefont {M.~P.}\ \bibnamefont {Waller}},\
  }\href@noop {} {\bibfield  {journal} {\bibinfo  {journal} {Nature}\ }\textbf
  {\bibinfo {volume} {555}},\ \bibinfo {pages} {604} (\bibinfo {year}
  {2018})}\BibitemShut {NoStop}%
\bibitem [{\citenamefont {Gao}\ \emph {et~al.}(2018)\citenamefont {Gao},
  \citenamefont {Struble}, \citenamefont {Coley}, \citenamefont {Wang},
  \citenamefont {Green},\ and\ \citenamefont {Jensen}}]{gao2018using}%
  \BibitemOpen
  \bibfield  {author} {\bibinfo {author} {\bibfnamefont {H.}~\bibnamefont
  {Gao}}, \bibinfo {author} {\bibfnamefont {T.~J.}\ \bibnamefont {Struble}},
  \bibinfo {author} {\bibfnamefont {C.~W.}\ \bibnamefont {Coley}}, \bibinfo
  {author} {\bibfnamefont {Y.}~\bibnamefont {Wang}}, \bibinfo {author}
  {\bibfnamefont {W.~H.}\ \bibnamefont {Green}}, \ and\ \bibinfo {author}
  {\bibfnamefont {K.~F.}\ \bibnamefont {Jensen}},\ }\href@noop {} {\bibfield
  {journal} {\bibinfo  {journal} {ACS Central Science}\ } (\bibinfo {year}
  {2018})}\BibitemShut {NoStop}%
\bibitem [{\citenamefont {Coley}\ \emph {et~al.}(2017)\citenamefont {Coley},
  \citenamefont {Barzilay}, \citenamefont {Jaakkola}, \citenamefont {Green},\
  and\ \citenamefont {Jensen}}]{coley2017prediction}%
  \BibitemOpen
  \bibfield  {author} {\bibinfo {author} {\bibfnamefont {C.~W.}\ \bibnamefont
  {Coley}}, \bibinfo {author} {\bibfnamefont {R.}~\bibnamefont {Barzilay}},
  \bibinfo {author} {\bibfnamefont {T.~S.}\ \bibnamefont {Jaakkola}}, \bibinfo
  {author} {\bibfnamefont {W.~H.}\ \bibnamefont {Green}}, \ and\ \bibinfo
  {author} {\bibfnamefont {K.~F.}\ \bibnamefont {Jensen}},\ }\href@noop {}
  {\bibfield  {journal} {\bibinfo  {journal} {ACS Central Science}\ }\textbf
  {\bibinfo {volume} {3}},\ \bibinfo {pages} {434} (\bibinfo {year}
  {2017})}\BibitemShut {NoStop}%
\bibitem [{\citenamefont {Saal}\ \emph {et~al.}(2013)\citenamefont {Saal},
  \citenamefont {Kirklin}, \citenamefont {Aykol}, \citenamefont {Meredig},\
  and\ \citenamefont {Wolverton}}]{saal2013materials}%
  \BibitemOpen
  \bibfield  {author} {\bibinfo {author} {\bibfnamefont {J.~E.}\ \bibnamefont
  {Saal}}, \bibinfo {author} {\bibfnamefont {S.}~\bibnamefont {Kirklin}},
  \bibinfo {author} {\bibfnamefont {M.}~\bibnamefont {Aykol}}, \bibinfo
  {author} {\bibfnamefont {B.}~\bibnamefont {Meredig}}, \ and\ \bibinfo
  {author} {\bibfnamefont {C.}~\bibnamefont {Wolverton}},\ }\href@noop {}
  {\bibfield  {journal} {\bibinfo  {journal} {Jom}\ }\textbf {\bibinfo {volume}
  {65}},\ \bibinfo {pages} {1501} (\bibinfo {year} {2013})}\BibitemShut
  {NoStop}%
\bibitem [{\citenamefont {Jain}\ \emph {et~al.}(2013)\citenamefont {Jain},
  \citenamefont {Ong}, \citenamefont {Hautier}, \citenamefont {Chen},
  \citenamefont {Richards}, \citenamefont {Dacek}, \citenamefont {Cholia},
  \citenamefont {Gunter}, \citenamefont {Skinner}, \citenamefont {Ceder} \emph
  {et~al.}}]{jain2013commentary}%
  \BibitemOpen
  \bibfield  {author} {\bibinfo {author} {\bibfnamefont {A.}~\bibnamefont
  {Jain}}, \bibinfo {author} {\bibfnamefont {S.~P.}\ \bibnamefont {Ong}},
  \bibinfo {author} {\bibfnamefont {G.}~\bibnamefont {Hautier}}, \bibinfo
  {author} {\bibfnamefont {W.}~\bibnamefont {Chen}}, \bibinfo {author}
  {\bibfnamefont {W.~D.}\ \bibnamefont {Richards}}, \bibinfo {author}
  {\bibfnamefont {S.}~\bibnamefont {Dacek}}, \bibinfo {author} {\bibfnamefont
  {S.}~\bibnamefont {Cholia}}, \bibinfo {author} {\bibfnamefont
  {D.}~\bibnamefont {Gunter}}, \bibinfo {author} {\bibfnamefont
  {D.}~\bibnamefont {Skinner}}, \bibinfo {author} {\bibfnamefont
  {G.}~\bibnamefont {Ceder}},  \emph {et~al.},\ }\href@noop {} {\bibfield
  {journal} {\bibinfo  {journal} {APL Materials}\ }\textbf {\bibinfo {volume}
  {1}},\ \bibinfo {pages} {011002} (\bibinfo {year} {2013})}\BibitemShut
  {NoStop}%
\bibitem [{\citenamefont {Goodman}(2009)}]{goodman2009computer}%
  \BibitemOpen
  \bibfield  {author} {\bibinfo {author} {\bibfnamefont {J.}~\bibnamefont
  {Goodman}},\ }\href@noop {} {\enquote {\bibinfo {title} {Computer software
  review: Reaxys},}\ } (\bibinfo {year} {2009})\BibitemShut {NoStop}%
\bibitem [{\citenamefont {Kim}\ \emph {et~al.}(2017{\natexlab{a}})\citenamefont
  {Kim}, \citenamefont {Huang}, \citenamefont {Saunders}, \citenamefont
  {McCallum}, \citenamefont {Ceder},\ and\ \citenamefont
  {Olivetti}}]{kim2017materials}%
  \BibitemOpen
  \bibfield  {author} {\bibinfo {author} {\bibfnamefont {E.}~\bibnamefont
  {Kim}}, \bibinfo {author} {\bibfnamefont {K.}~\bibnamefont {Huang}}, \bibinfo
  {author} {\bibfnamefont {A.}~\bibnamefont {Saunders}}, \bibinfo {author}
  {\bibfnamefont {A.}~\bibnamefont {McCallum}}, \bibinfo {author}
  {\bibfnamefont {G.}~\bibnamefont {Ceder}}, \ and\ \bibinfo {author}
  {\bibfnamefont {E.}~\bibnamefont {Olivetti}},\ }\href@noop {} {\bibfield
  {journal} {\bibinfo  {journal} {Chemistry of Materials}\ }\textbf {\bibinfo
  {volume} {29}},\ \bibinfo {pages} {9436} (\bibinfo {year}
  {2017}{\natexlab{a}})}\BibitemShut {NoStop}%
\bibitem [{\citenamefont {Raccuglia}\ \emph {et~al.}(2016)\citenamefont
  {Raccuglia}, \citenamefont {Elbert}, \citenamefont {Adler}, \citenamefont
  {Falk}, \citenamefont {Wenny}, \citenamefont {Mollo}, \citenamefont {Zeller},
  \citenamefont {Friedler}, \citenamefont {Schrier},\ and\ \citenamefont
  {Norquist}}]{raccuglia2016machine}%
  \BibitemOpen
  \bibfield  {author} {\bibinfo {author} {\bibfnamefont {P.}~\bibnamefont
  {Raccuglia}}, \bibinfo {author} {\bibfnamefont {K.~C.}\ \bibnamefont
  {Elbert}}, \bibinfo {author} {\bibfnamefont {P.~D.}\ \bibnamefont {Adler}},
  \bibinfo {author} {\bibfnamefont {C.}~\bibnamefont {Falk}}, \bibinfo {author}
  {\bibfnamefont {M.~B.}\ \bibnamefont {Wenny}}, \bibinfo {author}
  {\bibfnamefont {A.}~\bibnamefont {Mollo}}, \bibinfo {author} {\bibfnamefont
  {M.}~\bibnamefont {Zeller}}, \bibinfo {author} {\bibfnamefont {S.~A.}\
  \bibnamefont {Friedler}}, \bibinfo {author} {\bibfnamefont {J.}~\bibnamefont
  {Schrier}}, \ and\ \bibinfo {author} {\bibfnamefont {A.~J.}\ \bibnamefont
  {Norquist}},\ }\href@noop {} {\bibfield  {journal} {\bibinfo  {journal}
  {Nature}\ }\textbf {\bibinfo {volume} {533}},\ \bibinfo {pages} {73}
  (\bibinfo {year} {2016})}\BibitemShut {NoStop}%
\bibitem [{\citenamefont {Ghadbeigi}\ \emph {et~al.}(2015)\citenamefont
  {Ghadbeigi}, \citenamefont {Harada}, \citenamefont {Lettiere},\ and\
  \citenamefont {Sparks}}]{ghadbeigi2015performance}%
  \BibitemOpen
  \bibfield  {author} {\bibinfo {author} {\bibfnamefont {L.}~\bibnamefont
  {Ghadbeigi}}, \bibinfo {author} {\bibfnamefont {J.~K.}\ \bibnamefont
  {Harada}}, \bibinfo {author} {\bibfnamefont {B.~R.}\ \bibnamefont
  {Lettiere}}, \ and\ \bibinfo {author} {\bibfnamefont {T.~D.}\ \bibnamefont
  {Sparks}},\ }\href@noop {} {\bibfield  {journal} {\bibinfo  {journal} {Energy
  \& Environmental Science}\ }\textbf {\bibinfo {volume} {8}},\ \bibinfo
  {pages} {1640} (\bibinfo {year} {2015})}\BibitemShut {NoStop}%
\bibitem [{\citenamefont {Young}\ \emph {et~al.}(2018)\citenamefont {Young},
  \citenamefont {Maksov}, \citenamefont {Ziatdinov}, \citenamefont {Cao},
  \citenamefont {Burch}, \citenamefont {Balachandran}, \citenamefont {Li},
  \citenamefont {Somnath}, \citenamefont {Patton}, \citenamefont {Kalinin}
  \emph {et~al.}}]{young2018data}%
  \BibitemOpen
  \bibfield  {author} {\bibinfo {author} {\bibfnamefont {S.~R.}\ \bibnamefont
  {Young}}, \bibinfo {author} {\bibfnamefont {A.}~\bibnamefont {Maksov}},
  \bibinfo {author} {\bibfnamefont {M.}~\bibnamefont {Ziatdinov}}, \bibinfo
  {author} {\bibfnamefont {Y.}~\bibnamefont {Cao}}, \bibinfo {author}
  {\bibfnamefont {M.}~\bibnamefont {Burch}}, \bibinfo {author} {\bibfnamefont
  {J.}~\bibnamefont {Balachandran}}, \bibinfo {author} {\bibfnamefont
  {L.}~\bibnamefont {Li}}, \bibinfo {author} {\bibfnamefont {S.}~\bibnamefont
  {Somnath}}, \bibinfo {author} {\bibfnamefont {R.~M.}\ \bibnamefont {Patton}},
  \bibinfo {author} {\bibfnamefont {S.~V.}\ \bibnamefont {Kalinin}},  \emph
  {et~al.},\ }\href@noop {} {\bibfield  {journal} {\bibinfo  {journal} {Journal
  of Applied Physics}\ }\textbf {\bibinfo {volume} {123}},\ \bibinfo {pages}
  {115303} (\bibinfo {year} {2018})}\BibitemShut {NoStop}%
\bibitem [{\citenamefont {Kingma}\ and\ \citenamefont
  {Welling}(2014)}]{kingma2013auto}%
  \BibitemOpen
  \bibfield  {author} {\bibinfo {author} {\bibfnamefont {D.~P.}\ \bibnamefont
  {Kingma}}\ and\ \bibinfo {author} {\bibfnamefont {M.}~\bibnamefont
  {Welling}},\ }in\ \href@noop {} {\emph {\bibinfo {booktitle} {International
  Conference on Learning Representations}}}\ (\bibinfo {year}
  {2014})\BibitemShut {NoStop}%
\bibitem [{\citenamefont {Sohn}\ \emph {et~al.}(2015)\citenamefont {Sohn},
  \citenamefont {Lee},\ and\ \citenamefont {Yan}}]{sohn2015learning}%
  \BibitemOpen
  \bibfield  {author} {\bibinfo {author} {\bibfnamefont {K.}~\bibnamefont
  {Sohn}}, \bibinfo {author} {\bibfnamefont {H.}~\bibnamefont {Lee}}, \ and\
  \bibinfo {author} {\bibfnamefont {X.}~\bibnamefont {Yan}},\ }in\ \href@noop
  {} {\emph {\bibinfo {booktitle} {Advances in Neural Information Processing
  Systems}}}\ (\bibinfo {year} {2015})\ pp.\ \bibinfo {pages}
  {3483--3491}\BibitemShut {NoStop}%
\bibitem [{\citenamefont {G{\'o}mez-Bombarelli}\ \emph
  {et~al.}(2018)\citenamefont {G{\'o}mez-Bombarelli}, \citenamefont {Wei},
  \citenamefont {Duvenaud}, \citenamefont {Hern{\'a}ndez-Lobato}, \citenamefont
  {S{\'a}nchez-Lengeling}, \citenamefont {Sheberla}, \citenamefont
  {Aguilera-Iparraguirre}, \citenamefont {Hirzel}, \citenamefont {Adams},\ and\
  \citenamefont {Aspuru-Guzik}}]{gomez2018automatic}%
  \BibitemOpen
  \bibfield  {author} {\bibinfo {author} {\bibfnamefont {R.}~\bibnamefont
  {G{\'o}mez-Bombarelli}}, \bibinfo {author} {\bibfnamefont {J.~N.}\
  \bibnamefont {Wei}}, \bibinfo {author} {\bibfnamefont {D.}~\bibnamefont
  {Duvenaud}}, \bibinfo {author} {\bibfnamefont {J.~M.}\ \bibnamefont
  {Hern{\'a}ndez-Lobato}}, \bibinfo {author} {\bibfnamefont {B.}~\bibnamefont
  {S{\'a}nchez-Lengeling}}, \bibinfo {author} {\bibfnamefont {D.}~\bibnamefont
  {Sheberla}}, \bibinfo {author} {\bibfnamefont {J.}~\bibnamefont
  {Aguilera-Iparraguirre}}, \bibinfo {author} {\bibfnamefont {T.~D.}\
  \bibnamefont {Hirzel}}, \bibinfo {author} {\bibfnamefont {R.~P.}\
  \bibnamefont {Adams}}, \ and\ \bibinfo {author} {\bibfnamefont
  {A.}~\bibnamefont {Aspuru-Guzik}},\ }\href@noop {} {\bibfield  {journal}
  {\bibinfo  {journal} {ACS Central Science}\ }\textbf {\bibinfo {volume}
  {4}},\ \bibinfo {pages} {268} (\bibinfo {year} {2018})}\BibitemShut {NoStop}%
\bibitem [{Note1()}]{Note1}%
  \BibitemOpen
  \bibinfo {note} {\protect \url
  {www.github.com/olivettigroup/materials-synthesis-generative-models}}\BibitemShut
  {NoStop}%
\bibitem [{\citenamefont {Peters}\ \emph {et~al.}(2018)\citenamefont {Peters},
  \citenamefont {Neumann}, \citenamefont {Iyyer}, \citenamefont {Gardner},
  \citenamefont {Clark}, \citenamefont {Lee},\ and\ \citenamefont
  {Zettlemoyer}}]{peters2018deep}%
  \BibitemOpen
  \bibfield  {author} {\bibinfo {author} {\bibfnamefont {M.~E.}\ \bibnamefont
  {Peters}}, \bibinfo {author} {\bibfnamefont {M.}~\bibnamefont {Neumann}},
  \bibinfo {author} {\bibfnamefont {M.}~\bibnamefont {Iyyer}}, \bibinfo
  {author} {\bibfnamefont {M.}~\bibnamefont {Gardner}}, \bibinfo {author}
  {\bibfnamefont {C.}~\bibnamefont {Clark}}, \bibinfo {author} {\bibfnamefont
  {K.}~\bibnamefont {Lee}}, \ and\ \bibinfo {author} {\bibfnamefont
  {L.}~\bibnamefont {Zettlemoyer}},\ }in\ \href@noop {} {\emph {\bibinfo
  {booktitle} {Proceedings of the 2018 Conference of the North American Chapter
  of the Association for Computational Linguistics: Human Language
  Technologies, Volume 1 (Long Papers)}}}\ (\bibinfo {year} {2018})\BibitemShut
  {NoStop}%
\bibitem [{\citenamefont {Bojanowski}\ \emph {et~al.}(2017)\citenamefont
  {Bojanowski}, \citenamefont {Grave}, \citenamefont {Joulin},\ and\
  \citenamefont {Mikolov}}]{bojanowski2016enriching}%
  \BibitemOpen
  \bibfield  {author} {\bibinfo {author} {\bibfnamefont {P.}~\bibnamefont
  {Bojanowski}}, \bibinfo {author} {\bibfnamefont {E.}~\bibnamefont {Grave}},
  \bibinfo {author} {\bibfnamefont {A.}~\bibnamefont {Joulin}}, \ and\ \bibinfo
  {author} {\bibfnamefont {T.}~\bibnamefont {Mikolov}},\ }\href@noop {}
  {\bibfield  {journal} {\bibinfo  {journal} {Transactions of the Association
  for Computational Linguistics}\ }\textbf {\bibinfo {volume} {5}},\ \bibinfo
  {pages} {135} (\bibinfo {year} {2017})}\BibitemShut {NoStop}%
\bibitem [{\citenamefont {Kim}\ \emph {et~al.}(2017{\natexlab{b}})\citenamefont
  {Kim}, \citenamefont {Huang}, \citenamefont {Tomala}, \citenamefont
  {Matthews}, \citenamefont {Strubell}, \citenamefont {Saunders}, \citenamefont
  {McCallum},\ and\ \citenamefont {Olivetti}}]{kim2017machine}%
  \BibitemOpen
  \bibfield  {author} {\bibinfo {author} {\bibfnamefont {E.}~\bibnamefont
  {Kim}}, \bibinfo {author} {\bibfnamefont {K.}~\bibnamefont {Huang}}, \bibinfo
  {author} {\bibfnamefont {A.}~\bibnamefont {Tomala}}, \bibinfo {author}
  {\bibfnamefont {S.}~\bibnamefont {Matthews}}, \bibinfo {author}
  {\bibfnamefont {E.}~\bibnamefont {Strubell}}, \bibinfo {author}
  {\bibfnamefont {A.}~\bibnamefont {Saunders}}, \bibinfo {author}
  {\bibfnamefont {A.}~\bibnamefont {McCallum}}, \ and\ \bibinfo {author}
  {\bibfnamefont {E.}~\bibnamefont {Olivetti}},\ }\href@noop {} {\bibfield
  {journal} {\bibinfo  {journal} {Scientific Data}\ }\textbf {\bibinfo {volume}
  {4}},\ \bibinfo {pages} {170127} (\bibinfo {year}
  {2017}{\natexlab{b}})}\BibitemShut {NoStop}%
\bibitem [{\citenamefont {Chung}\ \emph {et~al.}(2014)\citenamefont {Chung},
  \citenamefont {Gulcehre}, \citenamefont {Cho},\ and\ \citenamefont
  {Bengio}}]{chung2014empirical}%
  \BibitemOpen
  \bibfield  {author} {\bibinfo {author} {\bibfnamefont {J.}~\bibnamefont
  {Chung}}, \bibinfo {author} {\bibfnamefont {C.}~\bibnamefont {Gulcehre}},
  \bibinfo {author} {\bibfnamefont {K.}~\bibnamefont {Cho}}, \ and\ \bibinfo
  {author} {\bibfnamefont {Y.}~\bibnamefont {Bengio}},\ }in\ \href@noop {}
  {\emph {\bibinfo {booktitle} {NIPS 2014 Workshop on Deep Learning, December
  2014}}}\ (\bibinfo {year} {2014})\BibitemShut {NoStop}%
\bibitem [{\citenamefont {Kamalakkannan}\ \emph {et~al.}(2016)\citenamefont
  {Kamalakkannan}, \citenamefont {Chandraboss},\ and\ \citenamefont
  {Senthilvelan}}]{kamalakkannan2016synthesis}%
  \BibitemOpen
  \bibfield  {author} {\bibinfo {author} {\bibfnamefont {J.}~\bibnamefont
  {Kamalakkannan}}, \bibinfo {author} {\bibfnamefont {V.}~\bibnamefont
  {Chandraboss}}, \ and\ \bibinfo {author} {\bibfnamefont {S.}~\bibnamefont
  {Senthilvelan}},\ }\href@noop {} {\bibfield  {journal} {\bibinfo  {journal}
  {World Scientific News}\ }\textbf {\bibinfo {volume} {58}},\ \bibinfo {pages}
  {97} (\bibinfo {year} {2016})}\BibitemShut {NoStop}%
\bibitem [{\citenamefont {Takatsu}\ \emph {et~al.}(2017)\citenamefont
  {Takatsu}, \citenamefont {Hernandez}, \citenamefont {Yoshimune},
  \citenamefont {Prestipino}, \citenamefont {Yamamoto}, \citenamefont {Tassel},
  \citenamefont {Kobayashi}, \citenamefont {Batuk}, \citenamefont {Shibata},
  \citenamefont {Abakumov} \emph {et~al.}}]{takatsu2017cubic}%
  \BibitemOpen
  \bibfield  {author} {\bibinfo {author} {\bibfnamefont {H.}~\bibnamefont
  {Takatsu}}, \bibinfo {author} {\bibfnamefont {O.}~\bibnamefont {Hernandez}},
  \bibinfo {author} {\bibfnamefont {W.}~\bibnamefont {Yoshimune}}, \bibinfo
  {author} {\bibfnamefont {C.}~\bibnamefont {Prestipino}}, \bibinfo {author}
  {\bibfnamefont {T.}~\bibnamefont {Yamamoto}}, \bibinfo {author}
  {\bibfnamefont {C.}~\bibnamefont {Tassel}}, \bibinfo {author} {\bibfnamefont
  {Y.}~\bibnamefont {Kobayashi}}, \bibinfo {author} {\bibfnamefont
  {D.}~\bibnamefont {Batuk}}, \bibinfo {author} {\bibfnamefont
  {Y.}~\bibnamefont {Shibata}}, \bibinfo {author} {\bibfnamefont {A.~M.}\
  \bibnamefont {Abakumov}},  \emph {et~al.},\ }\href@noop {} {\bibfield
  {journal} {\bibinfo  {journal} {Physical Review B}\ }\textbf {\bibinfo
  {volume} {95}},\ \bibinfo {pages} {155105} (\bibinfo {year}
  {2017})}\BibitemShut {NoStop}%
\bibitem [{\citenamefont {Sun}\ \emph {et~al.}(2017)\citenamefont {Sun},
  \citenamefont {Holder}, \citenamefont {Orva{\~n}anos}, \citenamefont {Arca},
  \citenamefont {Zakutayev}, \citenamefont {Lany},\ and\ \citenamefont
  {Ceder}}]{sun2017thermodynamic}%
  \BibitemOpen
  \bibfield  {author} {\bibinfo {author} {\bibfnamefont {W.}~\bibnamefont
  {Sun}}, \bibinfo {author} {\bibfnamefont {A.}~\bibnamefont {Holder}},
  \bibinfo {author} {\bibfnamefont {B.}~\bibnamefont {Orva{\~n}anos}}, \bibinfo
  {author} {\bibfnamefont {E.}~\bibnamefont {Arca}}, \bibinfo {author}
  {\bibfnamefont {A.}~\bibnamefont {Zakutayev}}, \bibinfo {author}
  {\bibfnamefont {S.}~\bibnamefont {Lany}}, \ and\ \bibinfo {author}
  {\bibfnamefont {G.}~\bibnamefont {Ceder}},\ }\href@noop {} {\bibfield
  {journal} {\bibinfo  {journal} {Chemistry of Materials}\ }\textbf {\bibinfo
  {volume} {29}},\ \bibinfo {pages} {6936} (\bibinfo {year}
  {2017})}\BibitemShut {NoStop}%
\bibitem [{\citenamefont {Hegde}\ \emph {et~al.}(2018)\citenamefont {Hegde},
  \citenamefont {Aykol}, \citenamefont {Kirklin},\ and\ \citenamefont
  {Wolverton}}]{hegde2018phase}%
  \BibitemOpen
  \bibfield  {author} {\bibinfo {author} {\bibfnamefont {V.~I.}\ \bibnamefont
  {Hegde}}, \bibinfo {author} {\bibfnamefont {M.}~\bibnamefont {Aykol}},
  \bibinfo {author} {\bibfnamefont {S.}~\bibnamefont {Kirklin}}, \ and\
  \bibinfo {author} {\bibfnamefont {C.}~\bibnamefont {Wolverton}},\ }\href@noop
  {} {\bibfield  {journal} {\bibinfo  {journal} {arXiv preprint
  arXiv:1808.10869}\ } (\bibinfo {year} {2018})}\BibitemShut {NoStop}%
\bibitem [{\citenamefont {Balachandran}\ \emph {et~al.}(2018)\citenamefont
  {Balachandran}, \citenamefont {Emery}, \citenamefont {Gubernatis},
  \citenamefont {Lookman}, \citenamefont {Wolverton},\ and\ \citenamefont
  {Zunger}}]{balachandran2018predictions}%
  \BibitemOpen
  \bibfield  {author} {\bibinfo {author} {\bibfnamefont {P.~V.}\ \bibnamefont
  {Balachandran}}, \bibinfo {author} {\bibfnamefont {A.~A.}\ \bibnamefont
  {Emery}}, \bibinfo {author} {\bibfnamefont {J.~E.}\ \bibnamefont
  {Gubernatis}}, \bibinfo {author} {\bibfnamefont {T.}~\bibnamefont {Lookman}},
  \bibinfo {author} {\bibfnamefont {C.}~\bibnamefont {Wolverton}}, \ and\
  \bibinfo {author} {\bibfnamefont {A.}~\bibnamefont {Zunger}},\ }\href@noop {}
  {\bibfield  {journal} {\bibinfo  {journal} {Physical Review Materials}\
  }\textbf {\bibinfo {volume} {2}},\ \bibinfo {pages} {043802} (\bibinfo {year}
  {2018})}\BibitemShut {NoStop}%
\bibitem [{\citenamefont {Yang}\ and\ \citenamefont
  {Ceder}(2013)}]{yang2013data}%
  \BibitemOpen
  \bibfield  {author} {\bibinfo {author} {\bibfnamefont {L.}~\bibnamefont
  {Yang}}\ and\ \bibinfo {author} {\bibfnamefont {G.}~\bibnamefont {Ceder}},\
  }\href@noop {} {\bibfield  {journal} {\bibinfo  {journal} {Physical Review
  B}\ }\textbf {\bibinfo {volume} {88}},\ \bibinfo {pages} {224107} (\bibinfo
  {year} {2013})}\BibitemShut {NoStop}%
\bibitem [{\citenamefont {Kusner}\ \emph {et~al.}(2017)\citenamefont {Kusner},
  \citenamefont {Paige},\ and\ \citenamefont
  {Hern{\'a}ndez-Lobato}}]{kusner2017grammar}%
  \BibitemOpen
  \bibfield  {author} {\bibinfo {author} {\bibfnamefont {M.~J.}\ \bibnamefont
  {Kusner}}, \bibinfo {author} {\bibfnamefont {B.}~\bibnamefont {Paige}}, \
  and\ \bibinfo {author} {\bibfnamefont {J.~M.}\ \bibnamefont
  {Hern{\'a}ndez-Lobato}},\ }in\ \href@noop {} {\emph {\bibinfo {booktitle}
  {International Conference on Machine Learning}}}\ (\bibinfo {year} {2017})\
  pp.\ \bibinfo {pages} {1945--1954}\BibitemShut {NoStop}%
\bibitem [{\citenamefont {Chollet}\ \emph {et~al.}(2015)\citenamefont {Chollet}
  \emph {et~al.}}]{chollet2015keras}%
  \BibitemOpen
  \bibfield  {author} {\bibinfo {author} {\bibfnamefont {F.}~\bibnamefont
  {Chollet}} \emph {et~al.},\ }\href@noop {} {\enquote {\bibinfo {title}
  {Keras},}\ } (\bibinfo {year} {2015})\BibitemShut {NoStop}%
\bibitem [{\citenamefont {Abadi}\ \emph {et~al.}(2016)\citenamefont {Abadi},
  \citenamefont {Barham}, \citenamefont {Chen}, \citenamefont {Chen},
  \citenamefont {Davis}, \citenamefont {Dean}, \citenamefont {Devin},
  \citenamefont {Ghemawat}, \citenamefont {Irving}, \citenamefont {Isard} \emph
  {et~al.}}]{abadi2016tensorflow}%
  \BibitemOpen
  \bibfield  {author} {\bibinfo {author} {\bibfnamefont {M.}~\bibnamefont
  {Abadi}}, \bibinfo {author} {\bibfnamefont {P.}~\bibnamefont {Barham}},
  \bibinfo {author} {\bibfnamefont {J.}~\bibnamefont {Chen}}, \bibinfo {author}
  {\bibfnamefont {Z.}~\bibnamefont {Chen}}, \bibinfo {author} {\bibfnamefont
  {A.}~\bibnamefont {Davis}}, \bibinfo {author} {\bibfnamefont
  {J.}~\bibnamefont {Dean}}, \bibinfo {author} {\bibfnamefont {M.}~\bibnamefont
  {Devin}}, \bibinfo {author} {\bibfnamefont {S.}~\bibnamefont {Ghemawat}},
  \bibinfo {author} {\bibfnamefont {G.}~\bibnamefont {Irving}}, \bibinfo
  {author} {\bibfnamefont {M.}~\bibnamefont {Isard}},  \emph {et~al.},\ }in\
  \href@noop {} {\emph {\bibinfo {booktitle} {OSDI}}},\ Vol.~\bibinfo {volume}
  {16}\ (\bibinfo {year} {2016})\ pp.\ \bibinfo {pages} {265--283}\BibitemShut
  {NoStop}%
\bibitem [{\citenamefont {Ong}\ \emph {et~al.}(2013)\citenamefont {Ong},
  \citenamefont {Richards}, \citenamefont {Jain}, \citenamefont {Hautier},
  \citenamefont {Kocher}, \citenamefont {Cholia}, \citenamefont {Gunter},
  \citenamefont {Chevrier}, \citenamefont {Persson},\ and\ \citenamefont
  {Ceder}}]{ong2013python}%
  \BibitemOpen
  \bibfield  {author} {\bibinfo {author} {\bibfnamefont {S.~P.}\ \bibnamefont
  {Ong}}, \bibinfo {author} {\bibfnamefont {W.~D.}\ \bibnamefont {Richards}},
  \bibinfo {author} {\bibfnamefont {A.}~\bibnamefont {Jain}}, \bibinfo {author}
  {\bibfnamefont {G.}~\bibnamefont {Hautier}}, \bibinfo {author} {\bibfnamefont
  {M.}~\bibnamefont {Kocher}}, \bibinfo {author} {\bibfnamefont
  {S.}~\bibnamefont {Cholia}}, \bibinfo {author} {\bibfnamefont
  {D.}~\bibnamefont {Gunter}}, \bibinfo {author} {\bibfnamefont {V.~L.}\
  \bibnamefont {Chevrier}}, \bibinfo {author} {\bibfnamefont {K.~A.}\
  \bibnamefont {Persson}}, \ and\ \bibinfo {author} {\bibfnamefont
  {G.}~\bibnamefont {Ceder}},\ }\href@noop {} {\bibfield  {journal} {\bibinfo
  {journal} {Computational Materials Science}\ }\textbf {\bibinfo {volume}
  {68}},\ \bibinfo {pages} {314} (\bibinfo {year} {2013})}\BibitemShut
  {NoStop}%
\bibitem [{\citenamefont {Swain}\ and\ \citenamefont
  {Cole}(2016)}]{swain2016chemdataextractor}%
  \BibitemOpen
  \bibfield  {author} {\bibinfo {author} {\bibfnamefont {M.~C.}\ \bibnamefont
  {Swain}}\ and\ \bibinfo {author} {\bibfnamefont {J.~M.}\ \bibnamefont
  {Cole}},\ }\href@noop {} {\bibfield  {journal} {\bibinfo  {journal} {Journal
  of chemical information and modeling}\ }\textbf {\bibinfo {volume} {56}},\
  \bibinfo {pages} {1894} (\bibinfo {year} {2016})}\BibitemShut {NoStop}%
\bibitem [{\citenamefont {Katyayan}\ and\ \citenamefont
  {Agrawal}(2017)}]{katyayan2017investigation}%
  \BibitemOpen
  \bibfield  {author} {\bibinfo {author} {\bibfnamefont {S.}~\bibnamefont
  {Katyayan}}\ and\ \bibinfo {author} {\bibfnamefont {S.}~\bibnamefont
  {Agrawal}},\ }\href@noop {} {\bibfield  {journal} {\bibinfo  {journal}
  {Journal of Materials Science: Materials in Electronics}\ }\textbf {\bibinfo
  {volume} {28}},\ \bibinfo {pages} {18442} (\bibinfo {year}
  {2017})}\BibitemShut {NoStop}%
\bibitem [{\citenamefont {Larsson}\ \emph {et~al.}(2008)\citenamefont
  {Larsson}, \citenamefont {Withers}, \citenamefont {Perez-Mato}, \citenamefont
  {Gerald}, \citenamefont {Saines}, \citenamefont {Kennedy},\ and\
  \citenamefont {Liu}}]{larsson2008microstructure}%
  \BibitemOpen
  \bibfield  {author} {\bibinfo {author} {\bibfnamefont {A.-K.}\ \bibnamefont
  {Larsson}}, \bibinfo {author} {\bibfnamefont {R.}~\bibnamefont {Withers}},
  \bibinfo {author} {\bibfnamefont {J.}~\bibnamefont {Perez-Mato}}, \bibinfo
  {author} {\bibfnamefont {J.~F.}\ \bibnamefont {Gerald}}, \bibinfo {author}
  {\bibfnamefont {P.~J.}\ \bibnamefont {Saines}}, \bibinfo {author}
  {\bibfnamefont {B.~J.}\ \bibnamefont {Kennedy}}, \ and\ \bibinfo {author}
  {\bibfnamefont {Y.}~\bibnamefont {Liu}},\ }\href@noop {} {\bibfield
  {journal} {\bibinfo  {journal} {Journal of Solid State Chemistry}\ }\textbf
  {\bibinfo {volume} {181}},\ \bibinfo {pages} {1816} (\bibinfo {year}
  {2008})}\BibitemShut {NoStop}%
\end{thebibliography}%
\end{document}